\documentclass[a4paper,10pt,twoside]{cpc-hepnp}

\usepackage{multicol}
\usepackage{graphicx}
\usepackage{amssymb,bm,mathrsfs,bbm,amscd}
\usepackage[tbtags]{amsmath}
\usepackage{lastpage}

\begin{document}

\title{Centrality Dependence of $\mathrm{{p}_{T}}$ Spectra for Identified Hadrons in
Au+Au and Cu+Cu Collisions at $\sqrt{s_{NN}}= 200$
GeV\thanks{Supported by the National Natural Science Foundation of
China (10775089, 10475049) and the science fund of Qufu Normal
University.} }

\author{%
Wang Yun-Fei$^{1}$
  \quad Shao Feng-Lan$^{1;1)}$\email{shaofl@mail.sdu.edu.cn}
  \quad Song Jun$^{1}$
  \quad Wei De-Ming$^{1}$
  \quad Xie Qu-Bing$^{2}$}

\maketitle

\address{
1~(Department of Physics, Qufu Normal University,
Qufu 273165, China)\\
2~(Department of Physics, Shandong University, Jinan 250100, China)\\
}

\begin{abstract}
The centrality dependence of transverse momentum spectra for
identified hadrons at midrapidity in Au+Au collisions at
$\sqrt{s_{NN}}= 200$ GeV is systematically studied in a quark
combination model. The $\mathrm{{p}_{T}}$ spectra of $\pi^{\pm}$,
$K^{\pm}$, $p(\bar{p})$ and $\Lambda(\bar{\Lambda})$ in different
centrality bins and the nuclear modification factors ($R_{CP}$) for
these hadrons are calculated. The centrality dependence of the
average collective transverse velocity $\langle \beta\, (r)\rangle$
for the hot and dense quark matter is obtained in Au+Au collisions,
and it is applied to a relative smaller Cu+Cu collision system. The
centrality dependence of $\mathrm{{p}_{T}}$ spectra and the $R_{CP}$
for $\pi^{0}$, $K_{s}^{0}$ and $\Lambda$ in Cu+Cu collisions at
$\sqrt{s_{NN}}= 200$ GeV are well described. The results show that
$\langle \beta\, (r)\rangle$ is only a function of the number of
participants $N_{part}$ and it is independent of the collision
system.
\end{abstract}



\footnotetext[0]{\hspace*{-2em}\small\centerline{\thepage\ --- \pageref{LastPage}}}%

\begin{multicols}{2}

\section{Introduction}

The hadronic collectivity is one of the important properties in
ultra-relativistic heavy ion collisions. It provides a lot of
information on the initial spatial anisotropy of the reaction
zone\cite{Reisdorf:1997,Sorge:1999,Kolb:1999,Teaney:1999}, the
degree of thermalization \cite{Kolb:2000,Voloshin:2000} and the
hadronization mechanism of the hot and dense medium produced in
nucleus-nucleus collisions\cite{Greco:2003prc,Fries:2003c}. It can
also help to understand the broadening of jetlike particle
correlations\cite{Armesto:2004} and high $\mathrm{{p}_{T}}$
jet-quenching\cite{XNWang:1992,Armesto:2005}. Strong partonic
multiple scatterings in nucleus-nucleus collisions would generate
the collectivity of quarks, which then result in the observed
collectivity of final hadrons. The quark number scaling of the
hadronic elliptic flow $\upsilon_{2}$ is a piece of evidence for
this original quark collectivity
\cite{Adare:2007,Adare:2003,Enokizono:2007}. Being a key
hadronization mechanism, quark combination picture has successfully
described many features of multi-particle production in high energy
heavy ion collisions, e.g. the high $p/\pi$ ratio in intermediate
transverse momentum
region\cite{Fries:2003,Greco:2003prl,Hwa:2003ppi}, the quark number
scaling behavior of hadron elliptic
flow\cite{Fries:2003c,Voloshin:2002wa,Molnar:2003ff} and its fine
structure at small $\mathrm{{p}_{T}}$\cite{Yao:2006fk}, hadron
longitudinal and transverse momentum
distributions\cite{Greco:2003prc,Hwa:2004,Shao:2007,Hwa:2005}, the
yields and multiplicity ratios\cite{Shao:2004cn}. Therefore, one can
extract the collectivity of the hot and dense quark matter from the
data of hadrons through quark combination mechanism.

In general, the nucleus-nucleus collisions in different centralities
will produce different sizes of the hot and dense quark matter. This
would cause the transverse collective flow for the hot and dense
quark matter varying with collision centralities. This variance of
collective flow in quark level would be embodied in the transverse
momentum spectra of final hadrons\cite{Arsene:2005,Adler:2004} and
particularly in their nuclear modification factors $R_{CP}$. The
thermal and hydrodynamic models have described the centrality
dependence of the transverse momentum distributions for final
hadrons in low $\mathrm{{p}_{T}}$ region with a statistical
hadronization
method\cite{Heinz:1992,Kolb:2003,Bass:2007,Heinz:2001,Huovinen:2005,Prorok:2006}.
Using the quark recombination at intermediate $\mathrm{{p}_{T}}$ and
parton fragmentation at high $\mathrm{{p}_{T}}$, the Duke group in
Ref.\ \cite{Fries:2003c} has explained the strong suppression of
hadron $R_{CP}$ at high transverse momenta, and the baryon-meson
difference of hadron $R_{CP}$ in intermediate $\mathrm{{p}_{T}}$
region. In the present paper, with quark combination at all
$\mathrm{{p}_{T}}$, we use our quark combination model to study the
hadron $\mathrm{{p}_{T}}$ spectra from central to peripheral
collisions. We investigate the $\mathrm{{p}_{T}}$ spectra of
identified hadrons at midrapidity in different centralities in Au+Au
collisions at $\sqrt{s_{NN}}= 200$ GeV to obtain the centrality
dependence of transverse collective flow for the hot and dense quark
matter. Furthermore, we apply it to relative smaller Cu+Cu collision
systems, calculate the transverse momentum spectra of final hadrons
and nuclear modification factors $R_{CP}$ in Cu+Cu collisions at
$\sqrt{s_{NN}}= 200$ GeV, and compare them with the experimental
data from STAR and PHENIX Collaborations.

In the next two sections, we briefly introduce the quark combination
model and the transverse momentum spectra of quarks just before
hadronization. The results and discussions are in Sect.~4. Summary
is given in Sect.~5.

\section{The quark combination model}

Within the same quark combination mechanism, all kinds of
combination-like models, such as recombination
model\cite{Fries:2003,Hwa:2004,Pratt:2005}, and coalescence
model\cite{Greco:2003prl,Molnar:2003ff}, have their own features.
Our quark combination model was first proposed for high energy
$e^+e^-$ and $pp$
collisions\cite{Xieqb:1988,Liang:1991ya,Wang:1995ch,
Zhao:1995hq,Wang:1996jy,Si:1997ux}. Recently we have extended the
model to ultra-relativistic heavy ion
collisions\cite{Yao:2006fk,Shao:2007,Shao:2004cn,Song:2007ph}. The
 model describes the production of initially
produced ground state mesons ($36-plets$) and baryons ($56-plets$).
In principle, it can also be applied to the production of excited
states \cite{Wang:1995ch} and exotic states\cite{Shao:2004cn}. These
hadrons through combination of constituent quarks are then allowed
to decay into the final state hadrons. We take into account the
decay contributions of all resonances of $56-plet$ baryons and
$36-plet$ mesons, and cover all available decay channels by using
the decay program of PYTHIA 6.1 \cite{Sjostrand}. The main idea is
to line up quarks and anti-quarks in a one-dimensional order in
phase space, e.g. in rapidity, and let them combine into initial
hadrons one by one following a combination rule. See the second
section of Ref. \cite{Shao:2004cn} for the short description of such
a rule. Of course, we also take into account the near correlation in
transverse momentum by limiting the maximum transverse momentum
difference for quarks and antiquarks as they combine into hadrons.
The flavor SU(3) symmetry with strangeness suppression in the yields
of initially produced hadrons is fulfilled in the model
\cite{Xieqb:1988,Wang:1995ch}.

\section{$\mathrm{{p}_{T}}$ spectra of the constituent quarks at hadronization}

It is known that the measured hadron $\mathrm{{p}_{T}}$ spectra in
relativistic heavy ion collisions exhibit a two-component behavior.
The spectra take an exponential form at low $\mathrm{{p}_{T}}$ and a
power-law form at high $\mathrm{{p}_{T}}$. Based on parton-hadron
duality, the transverse momentum spectra of constituent quarks just
before hadronization should also have the same property. In
principle, the quarks just before hadronization come from two parts,
i.e. the thermal quarks from the hot medium produced in collisions
and the minijet quarks from initial hard collisions. The final
hadrons are the total contribution of the two parts. But, just as
shown in the second figure in Ref.\ \cite{Fries:2003c}, the minijet
quarks dominate the large transverse momenta where thermal quarks
take a very small proportion, and thermal quarks dominate the low
transverse momentum region where the minijet quarks have a small
contribution. Therefore, we can neglect the two small contributions,
similar to the treatment in Ref.\cite{Greco:2003prl}, and adopt a
piecewise function to describe approximately the transverse momentum
distribution of constituent quarks:
\begin{eqnarray}
\dfrac{dN_{q}}{{2\pi\hspace{1mm}\mathrm{{p}_{T}}
 d\mathrm{\mathrm{{p}_{T}}}}}&=\theta(\mathrm{{p}_{0}}-\mathrm{{p}_{T}})N_{th}f_{th}(\mathrm{{p}_{T}})\nonumber\\
 &+
 \theta(\mathrm{{p}_{T}}-\mathrm{{p}_{0}})N_{jet}f_{jet}(\mathrm{{p}_{T}}),
 \label{quark-pt}
\end{eqnarray}
where $\theta(x)$ is the step function, $N_{th}$ is the number of
thermal quarks and $N_{jet}$ is the number of minijet quarks.
$\mathrm{{p}_{0}}$ is the transition point from thermal distribution
to power-law distribution, which is determined by the spectra
continuity. In fact, the interaction between the thermal quarks and
the minijet quarks leads to a smooth spectrum around
$\mathrm{{p}_{0}}$, and we neglect this effect in this paper.

The hot and dense quark matter produced in nucleus-nucleus
collisions at RHIC energies shows a significant collective
character\cite{Adare:2007,Adare:2003,Abelev:2007phi}. The
$\mathrm{{p}_{T}}$ spectra of thermal quarks at hadronization can be
described by a thermal phenomenological model incorporating the
transverse flow of thermal medium\cite{Heinz:1993}. The quarks and
antiquarks transversely boost with a flow velocity profile
$\beta_{r}(r)$ as a function of transverse radial position $r$.
$\beta_{r}(r)$ is parameterized by the surface velocity $\beta_{s}$:
$\beta_{r}(r)=\beta_{s}\,\xi^{\,n}$, where $\xi=r/R_{max}$, and
$R_{max}$ is the thermal source maximum radius ($0<\xi<1$). The
transverse flow of thermal medium can be equivalently described by a
superposition of a set of thermal sources, each boosted with
transverse rapidity $\rho=tanh^{-1}\beta_{r}$ \cite{Heinz:1993}:
\begin{eqnarray}
f_{th}(\mathrm{{p}_{T}})=\dfrac{dn_{th}}{{2\pi\hspace{1mm}\mathrm{{p}_{T}}
 d\mathrm{{p}_{T}}}}=&A\int_{0}^{1}\xi\,d\xi\,
 m_{T}\,
 \\ \nonumber \times{}
 I_{0}\bigg(\dfrac{\mathrm{{p}_{T}}\,sinh\,\rho}{T}\bigg)
&K_{1}\bigg( \dfrac{m_{T}\,cosh\,\rho}{T}\bigg),\nonumber
\label{thermal-pt}
\end{eqnarray}
where $A$ is the normalization constant in the region
$\mathrm{{p}_{T}}\in[0, \mathrm{{p}_{0}}]$. $I_{0}$ and $K_{1}$ are
the modified Bessel functions.
$m_T=\surd{\overline{{\mathrm{{p}_{T}}}^2+m^2}}$ is the transverse
mass of the
constituent quark. 
$T$ is the hadronization temperature. The average transverse
velocity can be written as
\begin{equation}
\langle \beta_{r}\rangle =\dfrac{\int\beta_{s}\, \xi^{\,n} \xi\,
d\xi}{\int \xi \,d\xi}=\dfrac{2}{n+2}\beta_{s}.
 \label{aver-beta}
\end{equation}
With fixed hadronization temperature $T=170$ MeV and parameter
$n=0.5$, the average transverse velocity $\langle \beta_r\rangle$ is
able to characterize the transverse collective flow of the hot and
dense quark matter.

The quarks and antiquarks with high transverse momenta are mainly
from the minijets created in initial hard collisions among nucleons.
Here the so-called minijet quarks are those just before
hadronization. They are the parton remnants after the revolution of
the initial hard partons by gluon radiation and split, and are
different from those in the fragmentation model. The
$\mathrm{{p}_{T}}$ spectra of minijet quarks at hadronization can be
parameterized as follows:
\begin{equation}
f_{jet}(\mathrm{{p}_{T}})=\dfrac{dn_{jet}}{{2\pi\hspace{1mm}\mathrm{{p}_{T}}
 d\mathrm{\mathrm{{p}_{T}}}}}=B\,\bigg(1+\dfrac{\mathrm{{p}_{T}}}{\mathrm{{p}_{0}}}\bigg)^{\,-\alpha},\\
\label{minijet-pt}
\end{equation}
where $B$ is the normalization constant in the region
$\mathrm{{p}_{T}}\in(\mathrm{{p}_{0}}, \infty )$.

There are four independent parameters in Eq.\ (\ref{quark-pt}) to
determine the transverse momentum distributions of quarks: $N_{th}$,
$N_{jet}$, $\langle \beta_r\rangle$ and $\alpha$. Here, we extract
the values of these parameters for the light and strange quark
$\mathrm{{p}_{T}}$ spectra at midrapidity from the data of $\pi^0$
and $K_s^0$\cite{Sakaguchi:2007,Adams:2006}, respectively. In our
quark combination model, removing the resonance decay contributions
from the measured $\pi^{0}$ and $K_s^0$ transverse momentum
distributions, we get the initially produced $\pi^{0}$ and $K_s^0$
transverse momentum spectra.  The values of parameters are inversely
extracted from these initial spectra. We obtain four groups of
results corresponding to the centrality bins $0-10\%$, $20-40\%$,
$40-60\%$, and $60-80\%$. They are shown in Fig.\ $1$. The lines in
the figure are the parameterized results, from which we can get the
quark $\mathrm{{p}_{T}}$ spectra in any collision centrality. In
Fig. 2, we also show the quark $\mathrm{{p}_{T}}$ spectra in four
collision centralities mentioned-above.

\begin{center}
\includegraphics[scale=0.40]{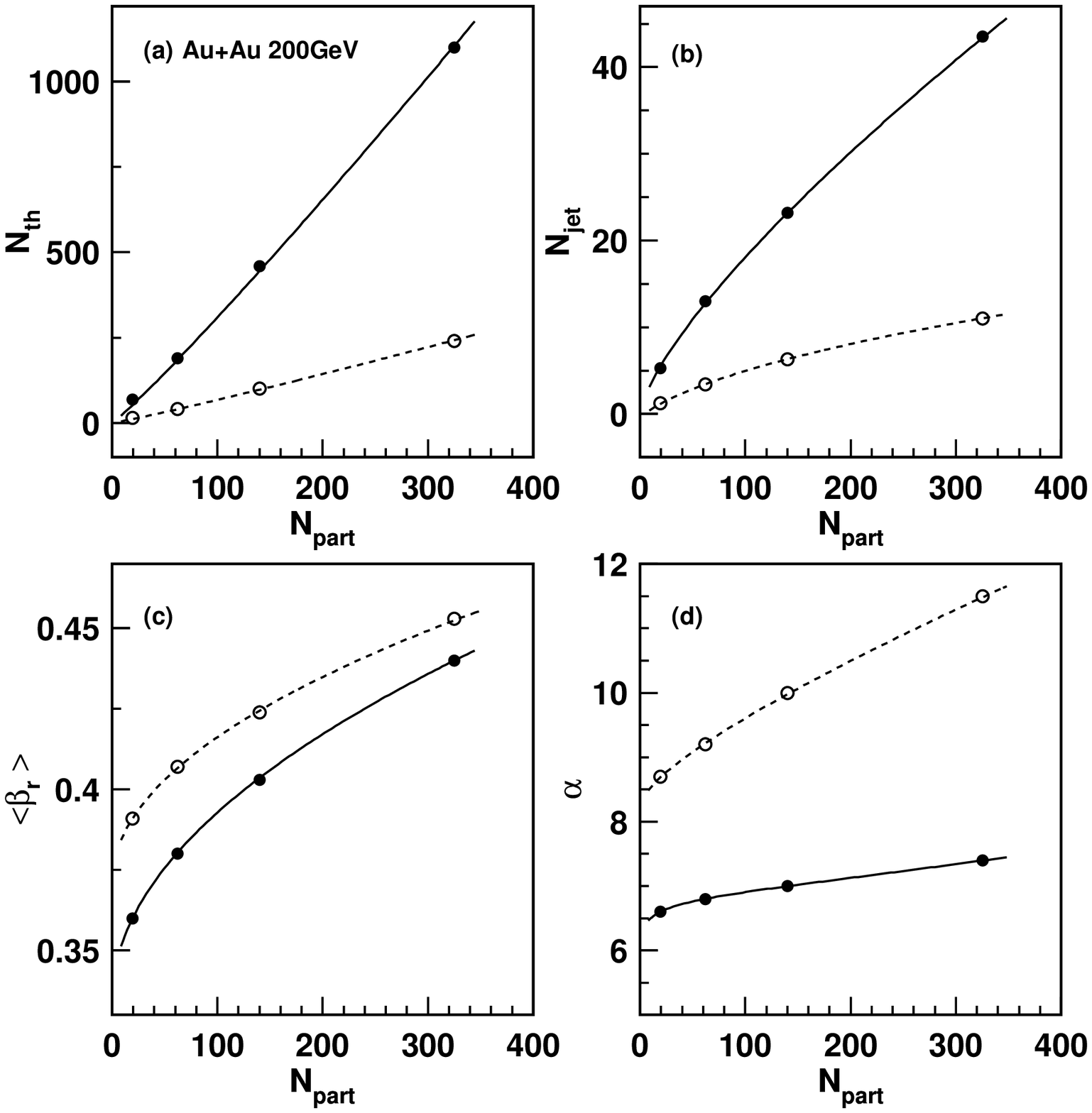}
\figcaption{\label{fig1}The values of the parameters $N_{th}$ (a),
$N_{jet}$ (b), $\langle \beta_r\rangle$ (c) and $\alpha$ (d) for the
$\mathrm{{p}_{T}}$ spectra of light quarks(filled circles) and
strange quarks(open circles) at midrapidity in four different
centrality bins. The corresponding centrality bins are $0-10\%$,
$20-40\%$, $40-60\%$, and $60-80\%$. The solid and dashed lines are
the parameterized results for light and strange quarks
respectively.}
\end{center}
\begin{center}
\vspace{-0.8cm}
\includegraphics[scale=0.4]{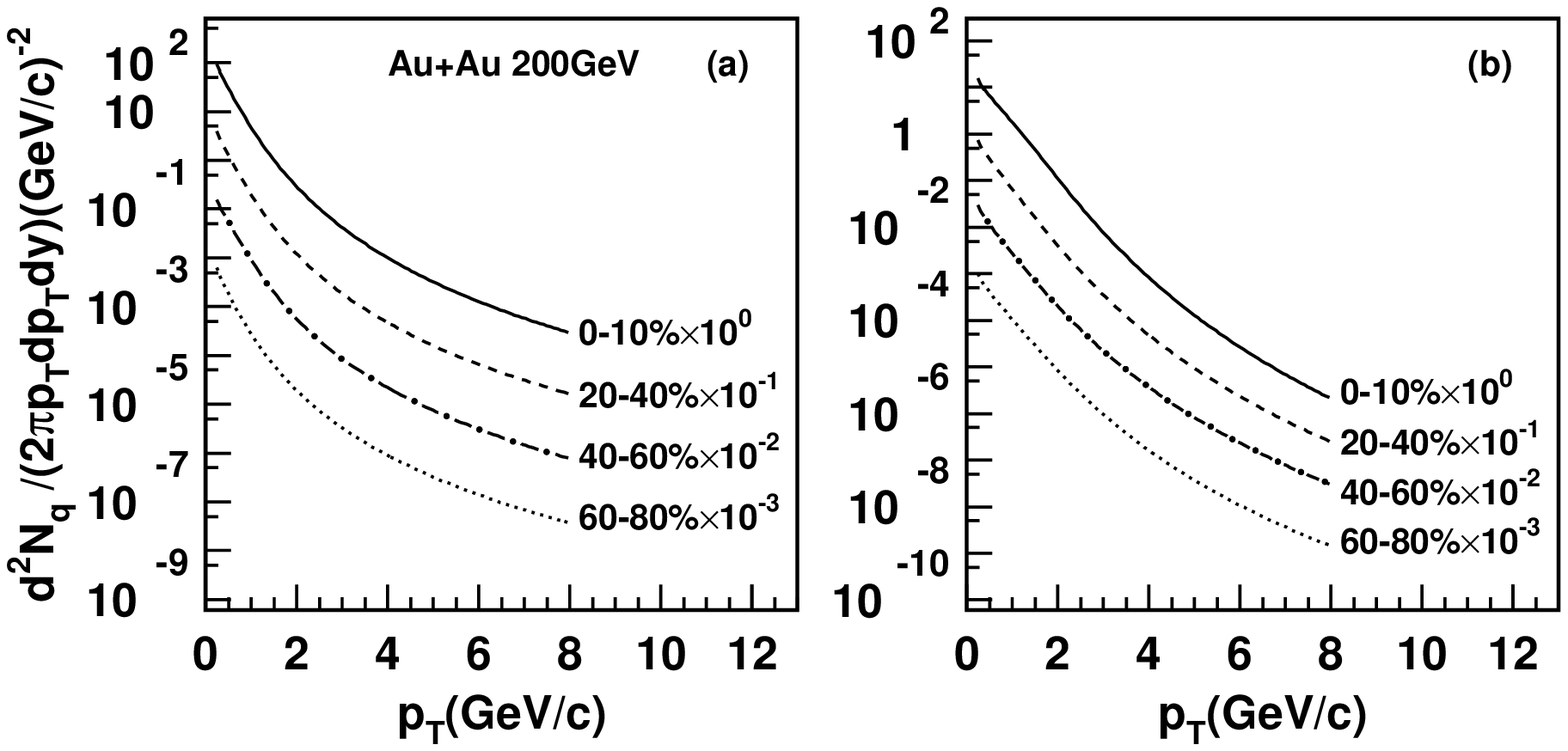}
\figcaption{The $\mathrm{{p}_{T}}$ spectra of the light quarks (a)
and strange quarks (b) at midrapidity in $Au+Au$ collisions at
$\sqrt{s_{NN}}= 200$ GeV.}
\end{center}

One can see from Fig. 1 that the centrality dependence of the
parameter $N_{th}$ is different from that of $N_{jet}$. As we know,
the thermal quarks which dominate the low $\mathrm{{p}_{T}}$ region
carry most of the collision energy. The available
 energy used to produce hadrons in different centrality is
proportional to the number of nucleon participants $N_{part}$ at the
fixed collision energy. Therefore, the number of thermal quarks is
approximately proportional to $N_{part}$. The minijet quarks with
high $\mathrm{{p}_{T}}$ are mainly from the hard-jet created in
initial hard collisions among nucleons, and their quantity is mainly
determined by the number of binary collisions $N_{coll}$, which is
obviously different from $N_{part}$. Generally, the higher the
collision centrality is, the bigger the bulk volume for hot medium,
the stronger the transverse collective flow $\beta$. The bigger the
bulk volume for hot medium is, the more energy loss for minijet
quarks, thus the steeper the minijet-quark spectra and the bigger
the parameter $\alpha$. It is further observed that the strange
quarks are different from light quarks not only in the quantity (due
to strangeness suppression) but also in the momentum distribution.
It is easy to understand that the spectrum of strange quarks at high
$\mathrm{{p}_{T}}$, due to its heavier effective mass, is steeper
than that of light quarks, i.e. $\alpha^{(s)}>\alpha^{(u,d)}$. In
the low $\mathrm{{p}_{T}}$ range, however, the spectrum of strange
quarks is flatter than the light quarks because
$\beta^{(s)}>\beta^{(u,d)}$. By analyzing the data of multi-strange
hadrons $\phi$, $\Xi$ and $\Omega$, Ref. \cite{Chen:2008} also draws
the same conclusion. Furthermore, the similar property is also
obtained in longitudinal orientation \cite{Song:2007}. As we know,
the expansion evolution of the plasma in the partonic phase is also
a process of obtaining the effective mass for partons. Due to the
heavier effective mass, the strange quarks may undergo a stronger
hydrodynamic expansion in the partonic phase than the light quarks.

As is known to all, though the nucleus-nucleus collisions at top
RHIC energy exhibit a high degree of transparency, there are still a
few net-quarks which stopped in the midrapidity region. The $p_T$
spectra of both $\pi^0$ and $K_s^0$ can not reflect the information
of net quarks\cite{Song:2007ph}. We obtain the number of net quarks
at midrapidity in different centrality bins by fitting the rapidity
densities of net-proton\cite{Adler:2004}. Note that the ratios of
$\pi^-/\pi^+$ and $\bar{p}/p$ measured by STAR and PHENIX
Collaborations reveal weak dependence of centrality and transverse
momentum\cite{Adler:2004,Abelev:2006}. The transverse momentum
distribution of net quarks is taken to be the same as that of the
newborn light quarks in the model.

With the input, we can give the transverse momentum distributions of
various hadrons in different collision centralities and the nuclear
modification factors $R_{CP}$ for these hadrons. Just as mentioned
in the above section, we consider the decay contributions from all
available decay channels of all resonances by using the decay
program of PYTHIA 6.1 \cite{Sjostrand}. Therefore, we can directly
compare our calculated results with the experimental data.

\section{Results and discussions}

\subsection{The $\mathrm{{p}_{T}}$ spectra of hadrons in Au+Au collisions}
We firstly calculate the transverse momentum spectra of $\pi^{\pm}$
and $p\, (\bar{p})$ in different centrality bins in Au+Au collisions
at $\sqrt{s_{NN}}= 200$ GeV. The results are shown in Fig.\ $2$.
Here, the pion spectra are corrected to remove the feed-down
contributions from $K_{s}^{0}$ and $\Lambda(\bar\Lambda)$. The
$\mathrm{{p}_{T}}$ spectra of hadrons in low $\mathrm{{p}_{T}}$
region are specially shown in the inserted plots. One can see that
the calculated results agree well with the data from the STAR
Collaboration. The strange hadron production can better reflect the
property of the hot and dense quark matter produced in collisions.
We also compute the transverse momentum distributions of strange
hadrons $K^{\pm}$, $\Lambda(\bar\Lambda)$ in different centrality
bins. The results are shown in Fig.\ $3$ and compared with the data.
\begin{center}
\vspace{-0.2cm}
\includegraphics[scale=0.24]{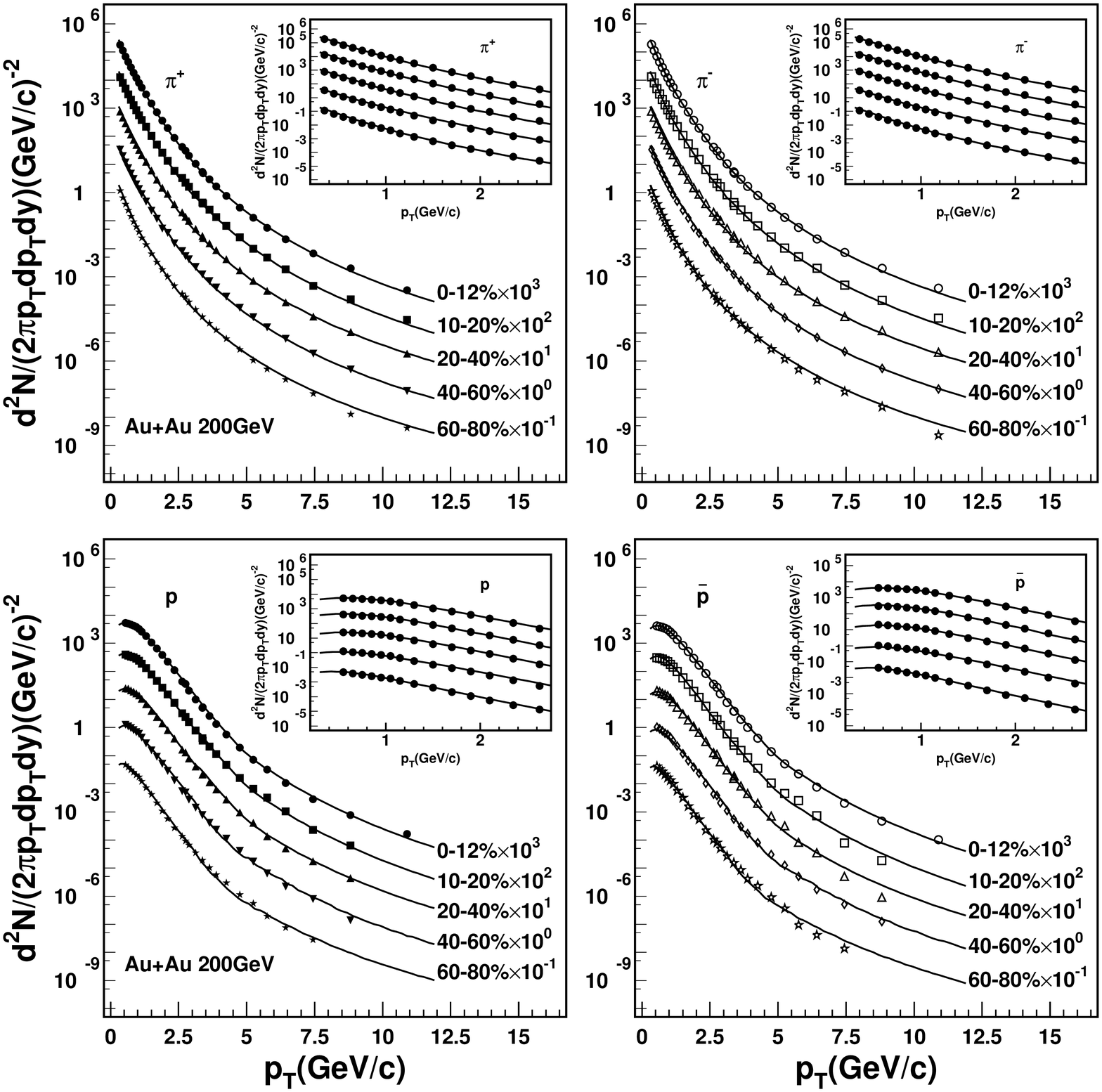}
\vspace{-0.4cm} \figcaption{\label{fig2}The transverse momentum
spectra of $\pi^{\pm}$ and $p\, (\bar{p})$ at midrapidity in
different centrality bins in Au+Au collisions at $\sqrt{s_{NN}}=
200$GeV. The data are taken from STAR
Collaboration~\citep{Abelev:2006}.}
\end{center}
\begin{center}
\vspace{-0.8cm}
\includegraphics[scale=0.24]{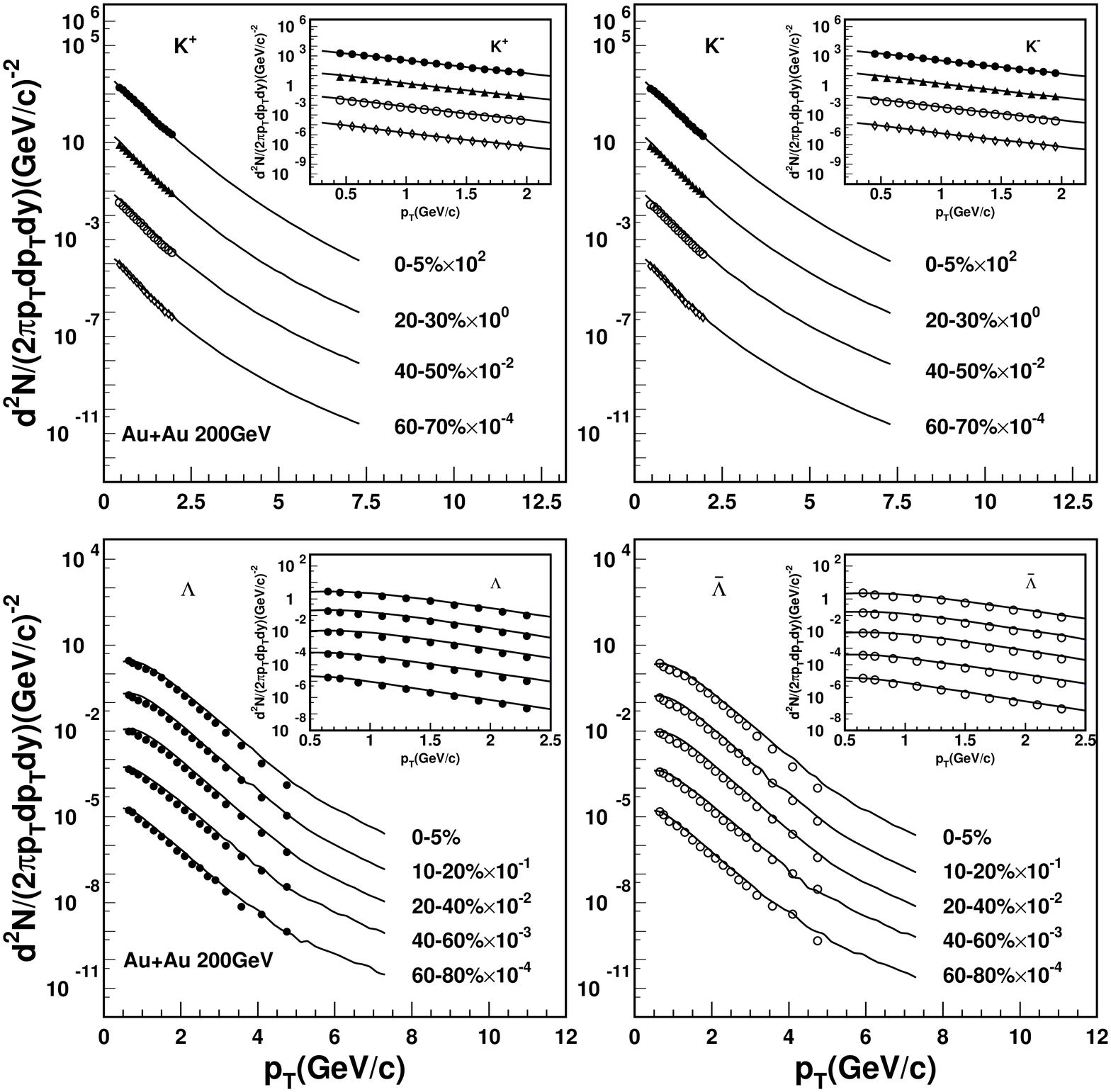}
\figcaption{\label{fig3}The transverse momentum spectra of strange
hadrons in different centrality bins in Au+Au collisions at
$\sqrt{s_{NN}}= 200$ GeV. The data are taken from PHENIX and STAR
Collaborations~\citep{Adler:2004,Admas:2006a}. }
\end{center}

As shown above, the good agreement between the model predictions and
the experimental data confirms the validity of our model from the
central to the peripheral collisions.

\subsection{The nuclear modification factors $R_{CP}$
 for hadrons in Au+Au collisions}

The nuclear modification factors $R_{CP}$ can reflect more precisely
the variation of hadron $\mathrm{{p}_{T}}$ spectra in different
centrality bins. It is quantified as\cite{Abelev:2006}:
\begin{equation}
\label{eqRcp} R_{CP}(\mathrm{{p}_{T}})=\frac{[(d^{2}N/(2\pi
\mathrm{{p}_{T}}
d\mathrm{{p}_{T}}dy))/N_{bin}]^{Central}}{[(d^{2}N/(2\pi
\mathrm{{p}_{T}} d\mathrm{{p}_{T}}dy))/N_{bin}]^{Peripheral}}.
\end{equation}
In Fig.\ $4$, we give the computed results of $R_{CP}$ for
$\pi^{+}+\pi^{-}$ and $p+\bar{p}$ ($0-10\%/60-80\%$), $K^{\pm}$ and
$\Lambda+\bar{\Lambda}$ (0-5\%/60-80\%), and compare them with the
experimental data.
\begin{center}
\includegraphics[scale=0.26]{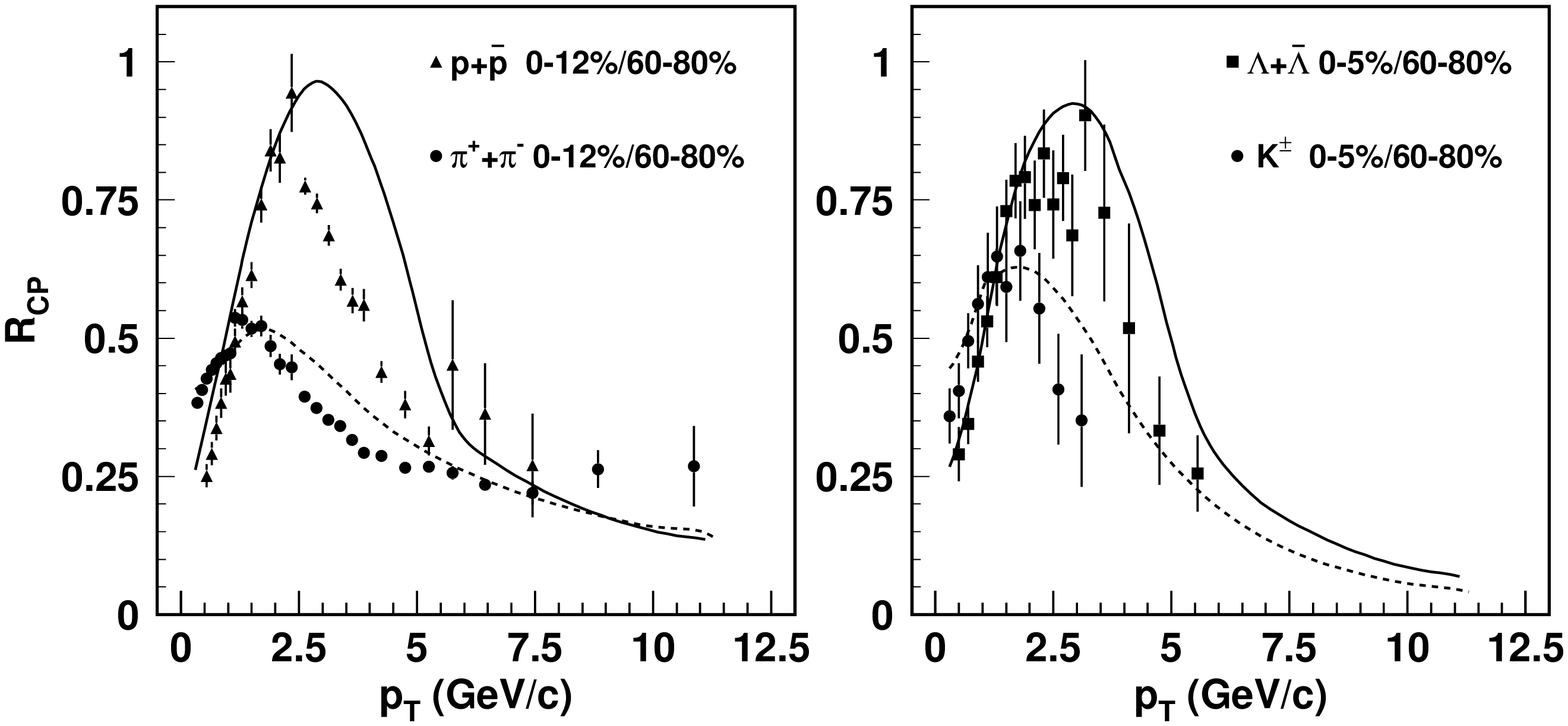}
\figcaption{\label{fig4}The nuclear modification factors $R_{CP}$
for identified hadrons in Au+Au collisions at $\sqrt{s_{NN}}= 200$
GeV. The data are taken from STAR
Collaboration~\citep{Abelev:2006,Admas:2006a,Adams:2004CP}.}
\end{center}

The hadrons in low $\mathrm{{p}_{T}}$ region are mainly from the
thermal quark combination. As shown in the first figure, the
transverse collective flow of the hot and dense quark matter,
denoted by $\langle \beta_r\rangle$, decreases with the falling
centrality. This leads to the $\mathrm{{p}_{T}}$ spectra of thermal
quarks becoming softer in peripheral collisions, and then results in
an increasing trend of $R_{CP}$ for hadrons in low
$\mathrm{{p}_{T}}$ region. The hadrons with intermediate
$\mathrm{{p}_{T}}$ are mainly produced by the combination of thermal
quarks with minijet quarks. As we all know, with the decreasing
collision centrality, the volume size of the hot and dense quark
matter becomes small and the energy loss of minijet quarks
correspondingly becomes small as they traverse the hot and dense
medium. It leads to the $\mathrm{{p}_{T}}$ spectra of minijet quarks
becoming harder in peripheral collisions, which is embodied in the
decreasing value of parameter $\alpha$ shown in the first figure
with the falling centrality. With the increasing of
$\mathrm{{p}_{T}}$, the hadron $R_{CP}$ goes over from the rise
caused by the combination of thermal quarks to fall caused by the
combination of minijet quarks. One can see that the quark
combination model well describes the behavior of $R_{CP}$ for final
hadrons in the whole $\mathrm{{p}_{T}}$ region considering the
variation of the transverse collective flow of the hot and dense
quark matter.

In addition, the data show that the $R_{CP}$ for baryons clearly
exhibits less suppression compared with that of mesons in
intermediate transverse momentum region. This type dependence of
$R_{CP}$, which is dependent upon the number of constituent quarks
rather than hadronic mass, has been qualitatively discussed earlier
in Refs.\ \cite{Fries:2003c,Adams:2004CP} as an experimental support
of the quark combination picture. Our results further manifest this
baryon-meson difference of $R_{CP}$ in intermediate
$\mathrm{{p}_{T}}$ region.

The experimental data show that the baryon-meson difference of
$R_{CP}$ disappears at higher $\mathrm{{p}_{T}}$. Using the
fragmentation mechanism, the Duke group in Ref.\ \cite{Fries:2003c}
has explained this common degree of suppression for both baryons and
mesons at high $\mathrm{{p}_{T}}$. Our results from the quark
combination can also describe this behavior of $R_{CP}$ at high
$\mathrm{{p}_{T}}$. These two different hadronization mechanisms
produce similar results. It suggests that the disappearance of
baryon-meson difference of $R_{CP}$ at high $\mathrm{{p}_{T}}$ is
not caused by hadronization mechanism. The hard scatterings which
take place near the surface of the collisions produce the
back-to-back dijets. One-side jets escape almost without energy
loss, while the away-side jets lose significant energy as they
traverse the hot and dense matter. The hadrons with high
$\mathrm{{p}_{T}}$ in all collision centralities are mostly from
these jets without energy loss. The transverse momentum
distributions of these minijet quarks with high $\mathrm{{p}_{T}}$
in central collisions, except the quantity, are almost the same with
those in peripheral collisions. Therefore, no matter what the
hadronization mechanism is, there exists a similar suppression of
the $R_{CP}$ for baryons and mesons at high $\mathrm{{p}_{T}}$.

It is also observed that the calculated $R_{CP}$ deviates from the
data to a certain degree. The reason may be that some effects, such
as the production of excited-state hadrons and final-state
rescattering, are not considered currently in the model. As we know,
a small quantity of the excited-state hadrons are produced in
relativistic heavy ion collisions, and the yields and momentum
distributions of final hadrons are influenced by the decay
contribution of these excited-state hadrons to a certain extent. As
the decay branch ratios of many excited-state hadrons are
incompletely measured, the contributions of excited-state hadrons
are neglected in the current model. On the other hand, the perfect
quark-number scaling of hadron elliptic flow $v_2$ suggests that the
influence of final-state rescattering on hadron distribution is
finite and small\cite{Adare:2007}, so we also neglect it in the
work. These two effects would affect the fine observable $R_{CP}$.
\subsection{The $\mathrm{{p}_{T}}$ spectra and nuclear modification
factors $R_{CP}$ for hadrons in Cu+Cu collisions}

Now, we apply the centrality(participants) dependence of parameters
in Au+Au collisions which determine the $\mathrm{{p}_{T}}$ spectra
of quarks in different centralities, to the relative smaller Cu+Cu
collision system at $\sqrt{s_{NN}}= 200$ GeV.  In Fig.\,$5$, we show
the calculation results for the transverse momentum spectra of
$\pi^{0}, K_{s}^{0}$ and hyperon $\Lambda$ in different centrality
bins and $R_{CP}$ for these hadrons in Cu+Cu collisions at
$\sqrt{s_{NN}}= 200$ GeV. The $\mathrm{{p}_{T}}$ spectra of hadrons
in low $\mathrm{{p}_{T}}$ region are specially shown in the inserted
plots. We find that the results agree well with the data in the low
transverse momentum region in all centrality bins. This implies that
the transverse collective flow of the hot and dense quark matter,
i.e. $\langle \beta_r\rangle$, is only a function of $N_{part}$ and
irrelevant to the collision system at the same collision energy. The
good agreement of our results with the $\pi^{0}$ data at high
transverse momenta also suggests that the energy loss of minijet
quarks in Cu+Cu collisions is almost the same with that in Au+Au
collisions with the same participants $N_{part}$ at $\sqrt{s_{NN}}=
200$ GeV. It is consistent with the recent measurement of STAR
Collaboration\cite{Hollis:2007,Selemon:2006}. The above results
suggest that the hot and dense quark matter produced in Au+Au and
Cu+Cu collisions at the same $N_{part}$ and collision energy has
similar strong-interacting character. Of course, even at the same
$N_{part}$ and collision energy, the initial spatial eccentricity of
the overlap collision geometry in Au+Au collisions is obviously
different from that in Cu+Cu collision systems. The difference is
clearly reflected by some important observations, e.g. the elliptic
flow of final hadrons and the global polarization of
hyperon\cite{Adare:2007,Adams:2004CP,Liang:2005prl,Abelev:2007polar,Back:2002}.
\end{multicols}
\begin{center}
\vspace{-0.3cm}
\includegraphics[scale=0.26]{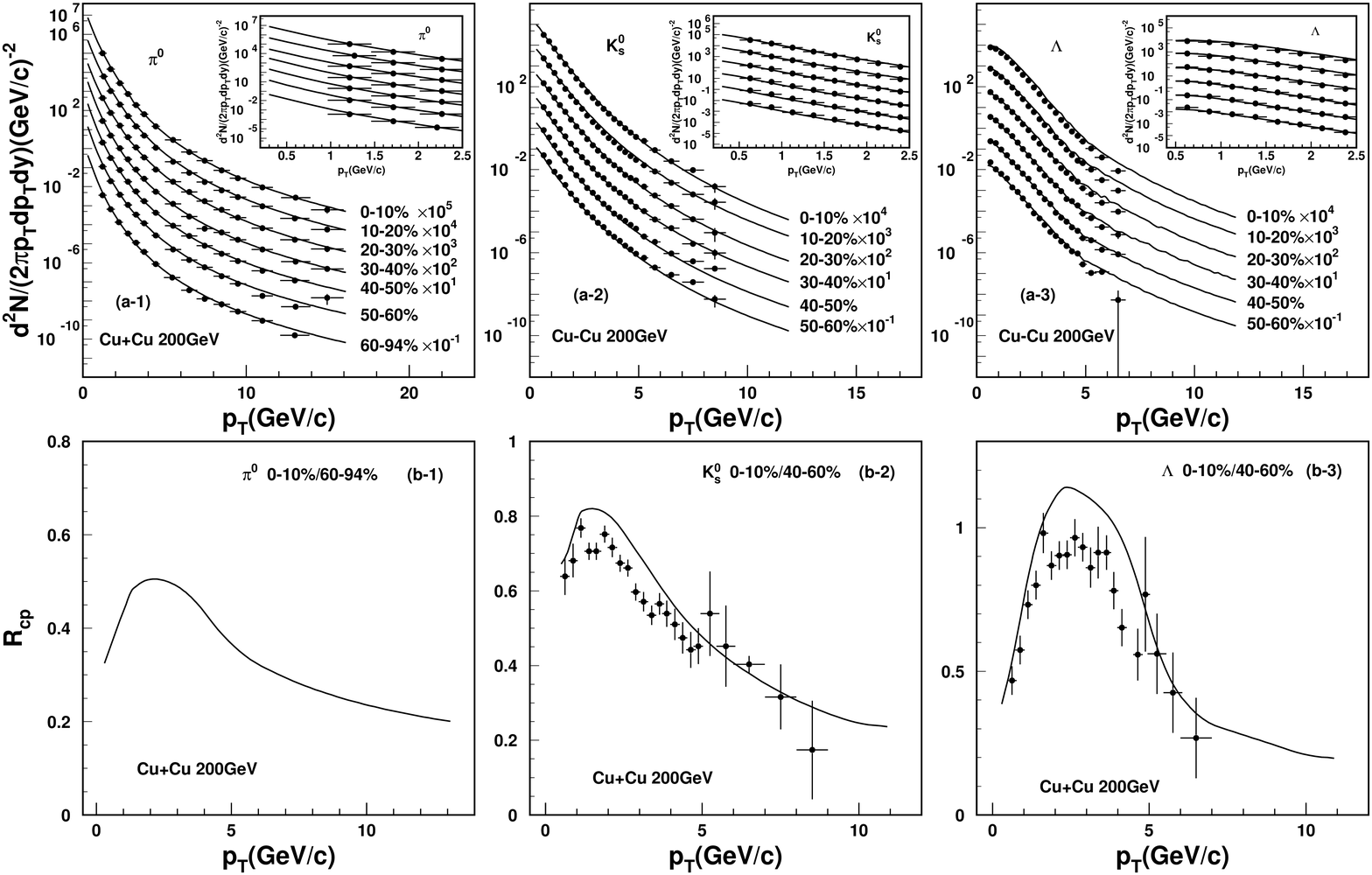}
\vspace{-0.3cm} \figcaption{\label{fig5}The transverse momentum
spectra of identified hadrons in different centrality bins (a) and
their nuclear modification factors $R_{CP}$ (b) in Cu+Cu collisions
at $\sqrt{s_{NN}}= 200$ GeV. The data are taken from PHENIX and STAR
Collaborations~\citep{Sakaguchi:2007,Timmins:2007}.}
\end{center}
\begin{multicols}{2}

\section{Summary}

Using the quark combination model, we study systematically the
centrality dependence of the transverse momentum distributions for
the identified hadrons at midrapidity in Au+Au and Cu+Cu collisions
at $\sqrt{s_{NN}}= 200$ GeV. The centrality dependence of the
parameters for quark $\mathrm{{p}_{T}}$ spectra is extracted from
the data of $\pi^{0}$ and $K_{s}^0$ in Au+Au collisions. We
calculate the transverse momentum distributions of $\pi^{\pm}$, $p\,
(\bar{p})$, $K^{\pm}$ and $\Lambda(\bar{\Lambda})$ in five
centrality bins. The good agreement between our results and the data
indicates that the quark combination hadronization mechanism is
applicable to all collision centralities. The nuclear modification
factors $R_{CP}$ for $\pi^{+}+\pi^{-}$ and $p+\bar{p}$
($0-10\%/60-80\%$), $K^{\pm}$ and
$\Lambda+\bar{\Lambda}$(0-5\%/60-80\%) are calculated and compared
with the data. The quark combination model well describes the
behavior of $R_{CP}$ for final hadrons in the whole
$\mathrm{{p}_{T}}$ region considering the decrease of the transverse
collective flow of the hot and dense quark matter from the central
collisions to the peripheral collisions. The disappearance of the
baryon-meson difference of $R_{CP}$ at higher $\mathrm{{p}_{T}}$ is
derived from the same transverse momentum distribution of minijet
quarks between the central and the peripheral collisions rather than
the hadronization mechanism.
 Furthermore, we apply the $N_{part}$ dependence of parameters
 to the relative smaller Cu+Cu collision system at the same collision
energy. We calculate the transverse momentum spectra of $\pi^{0}$,
$K_{s}^{0}$, and $\Lambda$ at midrapidity in different centrality
bins and $R_{CP}$ of these hadrons in Cu+Cu collisions at
$\sqrt{s_{NN}}= 200$ GeV. The results agree well with the data in
low transverse momentum region in all centrality bins. It suggests
that the transverse collective flow of the hot and dense quark
matter is only the function of $N_{part}$ and independent of
collision system. The calculated $\pi^{0}$ spectrum at high
$\mathrm{{p}_{T}}$ is also in good agreement with the data. These
results suggest that the hot and dense quark matter produced in
Au+Au and Cu+Cu collisions at the same $N_{part}$ and collision
energy has similar strong-interacting character.

\acknowledgments{We are grateful to Wang~Q.,~Liang~Z.~T.,~Yao~T. and
Han~W. for helpful discussions.}

\end{multicols}

\vspace{-2mm}
\centerline{\rule{80mm}{0.1pt}}
\vspace{2mm}

\begin{multicols}{2}

\end{multicols}

\clearpage

\end{document}